**Dissolution of donor-vacancy clusters in heavily doped n-type germanium**

Slawomir Prucnal[1], Maciej O. Liedke[2], Xiaoshuang Wang[1], Maik Butterling[2], Matthias Posselt[1], Joachim Knoch[3], Horst Windgassen[3], Eric Hirschmann[2], Yonder Berencén[1], Lars Rebohle[1], Mao Wang[1], Enrico Napoltani[4], Jacopo Frigerio[5], Andrea Ballabio[5], Giovani Isella[5], René Hübner[1], Andreas Wagner[2], Hartmut Bracht[6], Manfred Helm[1], and Shengqiang Zhou[1]

[1]*Helmholtz-Zentrum Dresden-Rossendorf, Institute of Ion Beam Physics and Materials Research, Bautzner Landstraße 400, D-01328 Dresden, Germany*

[2]*Helmholtz-Zentrum Dresden-Rossendorf, Institute of Radiation Physics, Bautzner Landstrasse 400, D-01328 Dresden, Germany*

[3]*Institut für Halbleitertechnik, RWTH Aachen, Germany*

[4]*Dipartimento di Fisica e Astronomia, Università di Padova and CNR-IMM MATIS, Via Marzolo 8, I-35131 Padova, Italy*

[5] *L-NESS, Dipartimento di Fisica, Politecnico di Milano, Polo di Como, Via Anzani 42, I-22100 Como, Italy*

[6] *Institute of Materials Physics, University of Münster, Wilhelm-Klemm-Str. 10, D-48149 Münster, Germany*

Corresponding author: Slawomir Prucnal, s.prucnal@hzdr.de

Abstract

The n-type doping of Ge is a self-limiting process due to the formation of vacancy-donor complexes ($D_nV$ with $n \leq 4$) that deactivate the donors. This work unambiguously demonstrates that the dissolution of the dominating $P_4V$ clusters in heavily phosphorus-doped Ge epilayers can be achieved by millisecond-flash lamp annealing at about 1050 K. The $P_4V$ cluster dissolution increases the carrier concentration by more than three-fold together with a suppression of phosphorus diffusion. Electrochemical capacitance-voltage measurements in conjunction with secondary ion mass spectrometry, positron annihilation lifetime spectroscopy and theoretical calculations enabled us to address and understand a fundamental problem that has hindered so far the full integration of Ge with complementary-metal-oxide-semiconductor technology.

1. **Introduction**

Germanium is the most Si-compatible high-mobility channel material. The electron and hole mobilities in Ge are about two and four times higher than in Si, respectively. Moreover, the carrier mobility in Ge can be easily enhanced by alloying Ge with Sn and by means of strain engineering [1-6]. While the fabrication of shallow junctions with highly-doped p-type Ge is well established, thermally stable n-type doping with a concentration of electrically active carriers above $5\times10^{19}$ cm$^{-3}$ remains challenging. In fact, using As implantation followed by laser melting or *in-situ* doping with Sb atoms allow the electron concentration reaching the level of $10^{20}$ cm$^{-3}$ [7, 8]. Unfortunately, due to the low solid solubility of As and Sb in Ge and their high diffusivity upon annealing such heavily doped layers are metastable [9, 10]. Milazzo *et al.* have shown that the deactivation of As in Ge starts already during annealing at 350 °C, a temperature used for the ohmic contact formation [7]. P having the solid solubility in Ge in the range of $2\times10^{20}$ cm$^{-3}$ is a shallow donor which is thermally stable during post-doping treatment and allows the effective doping above $10^{20}$ cm$^{-3}$. Shim *et al.* have shown that Ge heavily P-doped can be annealed up to 500 °C without diffusion [11]. In comparison to the main p-type dopant B, the preferred n-type dopants, namely P, As and Sb possess diffusivities several orders of magnitude higher [12-18]. The diffusion of the n-dopants that is controlled by foreign atoms occurs via the vacancy mechanism, i.e. via mobile dopant-vacancy pairs [17]. On the atomic level, this process is described by a ring-like mechanism. At high n-type dopant concentrations, large dopant-vacancy clusters can be formed [19, 20]. It has been shown by density functional theory (DFT) calculations, [21, 22] that these clusters are thermodynamically stable so that the number of electrically active monomers does not increase in the same manner as that of the n-type dopant atoms. The cluster formation is the reason why electrical activation of P and As dopants in Ge is much more difficult than that of B whose interaction with the vacancies is reported to be repulsive [17, 19, 20, 23-25].

This work provides an unambiguous experimental demonstration of the underlying mechanism enabling the increase of donor activation in heavily n-type doped Ge. As a model material, we have chosen ex-situ and in-situ P-doped layers of Ge epitaxially grown on Si wafer. Electrochemical capacitance-voltage (ECV) measurements in combination with secondary ion mass spectrometry (SIMS), positron annihilation spectroscopy (PAS) and theoretical calculations reveal that 20-millisecond-flash-lamp annealing (FLA) at 1050 ± 50 K provides enough energy to dissolve pre-existing phosphorus-vacancy clusters with an annealing time that is too short for a significant diffusion of phosphorus dopants. A part of the liberated monovacancies is trapped by existing larger defects or tends to form new small vacancy clusters.

## 2. Material and methods

### 2.1. Doping and annealing

We have investigated two types of Ge layers doped either in-situ or ex-situ using ion implantation followed by FLA. The in-situ doped, about 500-nm-thick Ge layers were grown epitaxially on (100) Si substrates by low-energy plasma-enhanced chemical vapor deposition (LEPECVD). Further details about the LEPECVD and the deposition of highly doped Ge epilayers can be found elsewhere [26, 27]. Basically, the single-crystalline Ge epilayers were grown at 773 K at a rate of ~1 nm/s using $GeH_4$ as the precursor. N-type doping of Ge was realized by adding $PH_3$ molecules into the reactor chamber with a $PH_3/GeH_4$ ratio of 1.7 %. In the as-grown layer, the maximum effective carrier concentration estimated from Hall Effect is found to be around $3.5\times10^{19}$ cm$^{-3}$, while the P distribution obtained by secondary ion mass spectrometry (SIMS) reveals that the concentration of P incorporated into Ge is in the range of $1\times10^{20}$ cm$^{-3}$. For ex-situ doping, about 400 nm-thick intrinsic-Ge layer was grown on Si substrate by molecular beam epitaxy (MBE). P was implanted with three different fluences ($1\times10^{15}$ cm$^{-2}$, $0.4\times10^{15}$ cm$^{-2}$, $4\times10^{15}$ cm$^{-2}$) and energies (60 keV, 100 keV, 180 keV) respectively in order to ensure the formation of about 300 nm-thick Ge doped layer with a P concentration of around $1.8\times10^{20}$ cm$^{-3}$. This implantation with three different fluences and energies allows to obtain a box-like distribution of P in Ge. After depositing the crystalline Ge layer, millisecond-FLA was performed. Ge samples were flash-lamp-annealed either from the front-side (f-FLA) or rear-side (r-FLA) for 20 ms at a maximum temperature of about 1050 ± 50 K in the continuous flow of nitrogen [28]. Such temperature is achieved during the flash with the energy density of 100 Jcm$^{-2}$. The increase of the flash energy density above 105 Jcm$^{-2}$ causes the melting of Ge layer, which should be avoided due to strong dopants segregation when Ge recrystallizes from liquid phase. In our FLA system (FLA100), an eight-inch wafer can be homogeneously annealed during a single flash with a maximum temperature easily exceeding the melting point of either Ge or Si. The maximum achieved temperature during the FLA process is self-limited by the melting temperature of the material to be annealed. The FLA for the current experiments was performed with the bank of twelve 26-cm-long Xe lamps with an emission spectrum covering the spectral range from UV to near infrared. During r-FLA, the flash-light is absorbed by the Si substrate and the heat wave propagates from Si to the Ge layer. In such a case, the Ge/Si interface is hotter than the Ge surface. In contrast, during (f-FLA), the main part of the flash spectrum is absorbed within the first 100 nm of the Ge layer. Therefore, during f-FLA, the Ge surface is the hottest. Those two different annealing procedures generate a temperature gradient across the Ge layer, either from the surface to the Ge/Si interface or in the opposite direction. The thermal stability of dopants were investigated using conventional rapid thermal annealing system. Samples were annealed for 100 s in continuous flow of $N_2$. The annealing temperature was in the range of 250 to 610 °C.

**2.2. Positron annihilation spectroscopy**

The type and distribution of defects in P-doped Ge were investigated by positron annihilation lifetime spectroscopy (PALS). The positron lifetime experiments were performed at the mono-energetic positron spectroscopy (MePS) beamline, which is a beamline of the radiation source ELBE (Electron Linac for beams with high Brilliance and low Emittance) at Helmholtz-Zentrum Dresden-Rossendorf (Germany) [29-31]. A digital lifetime $CrBr_3$ scintillator detector operated by a homemade software utilizing a SPDevices ADQ14DC-2X with 14-bit vertical resolution and 2GS/s horizontal resolution was used. This detector is optimized for room-temperature measurements and possesses a time resolution down to about 0.205 ns. The resolution function required for spectrum analysis is composed of two Gaussian functions with distinct intensities depending on the positron implantation energy, $E_p$, and appropriate relative shifts. All spectra contain at least $5 \times 10^6$ counts. The lifetime spectra were analyzed as a sum of time-dependent exponential decays (according to the equation: $N(t) = \Sigma_i I_i/\tau_i \cdot exp(-t/\tau_i)$) convoluted with the spectrometer timing resolution [32] using the non-linearly least-squared-based package PALSfit fitting software [33]. The indices $i$ corresponds to different discrete lifetime components in the spectra with individual lifetimes $\tau_i$ and intensities $I_i$. Regarding the accuracy of PALS, a spectrum with a single lifetime component or a spectrum with multiple components but sufficiently separated from each other can be fitted giving the accuracy down to 1 ps. In general, the possibility of obtaining a reliable multi-component fit is restricted due to the limited time resolution of the spectrometer and the statistics of the measurements. More technical details can be found in Ref. 34.

**Electrical and structural properties**

The concentration and the depth distribution of charge carriers in P-doped Ge layers were determined by ECV measurements. The ECV measurements were carried out at room temperature using EN.18 (EDTA 0.1m + NaOH 0.18m) electrolyte for etching. The total carrier concentration in the as-grown and flash-lamp-annealed samples was independently determined by Hall Effect measurements using a commercial Lakeshore Hall System with the van der Pauw geometry. The electron concentration in the in-situ doped samples increases after FLA from about $3.5 \times 10^{19}$ cm$^{-3}$ to $8.0 \times 10^{19}$ cm$^{-3}$, while the maximum electron concentration in the implanted and annealed samples was in the order of $1.2\times10^{20}$ cm$^{-3}$ (67 % of implanted P atoms are electrically active).

The Ge/Si interface quality was inspected using cross-sectional bright-field transmission electron microscopy (TEM) employing an image $C_s$-corrected Titan 80-300 (FEI) microscope operated at an accelerating voltage of 300 kV. In-depth P concentrations were measured by a dynamic secondary ion mass spectrometer (SIMS) with 3 keV $O_2^+$ sputter beam. A beam of $O_2^+$ ions was rastered over a surface area of $250\times250$ μm$^2$ and secondary ions were collected from the central part of the sputtered crater. Crater depths were measured with a Tencor P10 stylus profilometer, and a constant erosion rate was assumed when converting sputtering time to sample depth. The calibration of the P concentration was performed by measuring a Ge standard sample with known P areal density with an accuracy of ±10%.

### 3. Results and discussion
#### 3.1. Electrical and structural properties of P-doped Ge

Figure 1 shows the SIMS profile of the P distribution and the depth profiles of electrically active carriers measured by ECV. The atomic concentration of P is in the range of $1\times10^{20}$ cm$^{-3}$ and does not alter upon r-FLA or f-FLA. Obviously, the FLA process at about 1050 K for 20 ms is too short to cause significant P diffusion.

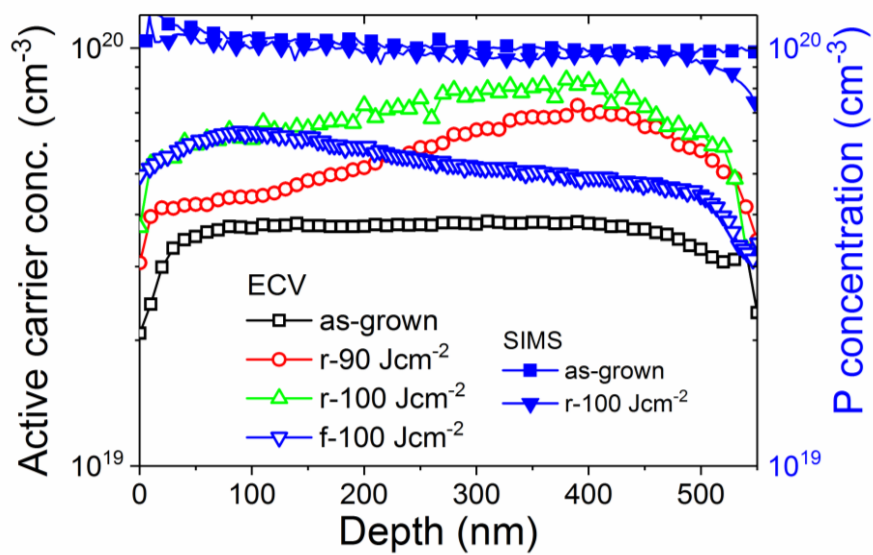

Figure 1. The depth distributions of electrically active carriers and P in Ge before and after FLA. The carrier distribution was obtained by ECV, while the P distribution was measured by SIMS. Samples was annealed either from the rear side (r-90 and r-100 Jcm$^{-2}$) or from the front side (f-100 Jcm$^{-2}$).

However, the comparison between the carrier concentration before and after annealing shows that the energy deposited by 20 ms-FLA is sufficient to increase the amount of electrically active P atoms from $3.5\times10^{19}$ cm$^{-3}$ to about $8.0\times10^{19}$ cm$^{-3}$ after r-FLA and to about $6.2\times10^{19}$ cm$^{-3}$ after f-FLA. In the as-grown sample, the carrier concentration is evenly distributed over the layer thickness. A significant enhancement of the activation efficiency of P in Ge was obtained after annealing with the FLA energy densities in the order of 90 Jcm$^{-2}$ or higher. The annealing of the sample with energy densities lower than 60 Jcm$^{-2}$ did not improve the carrier concentration. On the other hand, the annealing with the energy densities in the range of 60 - 90 Jcm$^{-2}$ slightly increases the carrier concentration. After r-FLA, the highest carrier concentration is detected about 100 nm above the Ge/Si interface and decreases slightly towards the surface. The profile of the carrier distribution follows the heat-wave propagation in the annealed sample from the Ge/Si interface towards the surface. In order to clarify the influence of the temperature gradient on the carrier distributions in P-doped Ge, the ECV measurements were also performed on samples annealed from the front side. After f-FLA the highest carrier concentration is close to the sample surface and decreases with depth as expected.

After r-FLA, the highest carrier concentration was found to be in a depth of about 100 nm above the Ge/Si interface, i.e. above the Ge layer region containing the highest concentration of structural defects (e.g. threading dislocations). The interface quality and the threading dislocation density in the as-grown and the annealed samples were investigated by cross-sectional TEM analysis. Figure 2 shows representative bright-field TEM images obtained from the as-grown sample (Fig. 2a) as well as from the samples flash-lamp-annealed from the rear side (Fig. 2b).

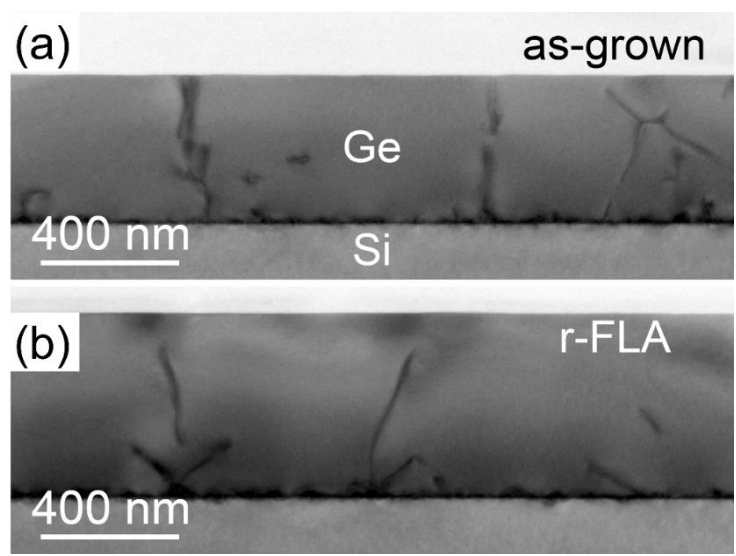

Figure 2 Representative bright-field TEM images of the as-grown (a) and r-FLA-treated sample (b).

We found a relatively high density of different structural defects at the Ge-Si interface as expected. These defects are created due to the lattice mismatch between Ge and Si. The treading dislocations and other large structural defects are effective trapping centres for donors [35]. Therefore, near the Ge/Si interface the effective carrier concentration is lower than in the centre of the film. During r-FLA, the heat wave penetrates the Ge layer from the Si substrate, which means that the thermal load in the vicinity of the Ge/Si interface region is the highest. In such a case, the highest carrier concentration was found to be about 100 nm above the interface region (above $8 \times 10^{19}$ cm$^{-3}$), since the highest thermal load is therein (see Fig. 1). Assuming that $P_xV$ clusters, where x≤4, are mainly responsible for the deactivation of P in Ge [36, 37], the dissociation of these clusters must take place during the millisecond-range annealing. The dissociation of $P_xV$ clusters is most pronounced at the depths where the thermal load is the highest. However, near the Ge/Si interface the density of dopant-vacancy clusters is higher than in other regions. Therefore, the carrier concentration just at the Ge/Si interface is a bit lower than in the center of the film.

We have also investigated the thermal stability of heavily doped n-type Ge using post-grown annealing. The P-doped Ge samples presented in figure 3 are samples fabricated either by LEPCVD (black squares) or ion implantation (red circles) while the As-doped Ge (green triangles) are data taken from Ref. 7. Before thermal stability test the P-doped samples were annealed from rear-side by FLA for 20 ms at 100 Jcm$^{-2}$. The deactivation of P was performed for 100 s in the temperature range of 250 to 610 °C. According to Re. 7 the As doped samples were annealed for 10 min. The deactivation of As doped Ge starts already at 250 °C. The effective carrier concentration reduces from $1\times10^{20}$ cm$^{-3}$ to $9\times10^{19}$ cm$^{-3}$. The increase of the annealing temperature up to 350 °C reduces the electron concentration down to $4.5\times10^{19}$ cm$^{-3}$ and after annealing at 550 °C the electron concentration decreases to $1.7\times10^{19}$ cm$^{-3}$, which is the solid solubility of As in Ge. Both P doped samples behave slightly different compared to As doped Ge. First the effective carrier concentration slightly increases after annealing at 300 °C and next it starts to decrease. The significant change of the effective carrier concentration was visible after annealing at the temperature higher than 550°C which is much more than for As doped samples. After annealing at 610 °C the effective carrier concentration in the in-situ doped and in ion implanted samples decreases to about $4\times10^{19}$ cm$^{-3}$ and $8.5\times10^{19}$ cm$^{-3}$. Much higher thermal stability of P-doped Ge can be due to higher activation enthalpy for diffusion of P than for As (2.85 eV for P vs. 2.71 eV for As) [13]. The activation enthalpy for Sb diffusion in Ge is even lower (2.55 eV) suggesting the lowest thermal stability. In Ge technology most of annealing steps, besides dopant activation, do not excess 500 °C. This means that P-implanted Ge can be effectively used as n-type channel material for Ge-based nanoelectronics and optoelectronics.

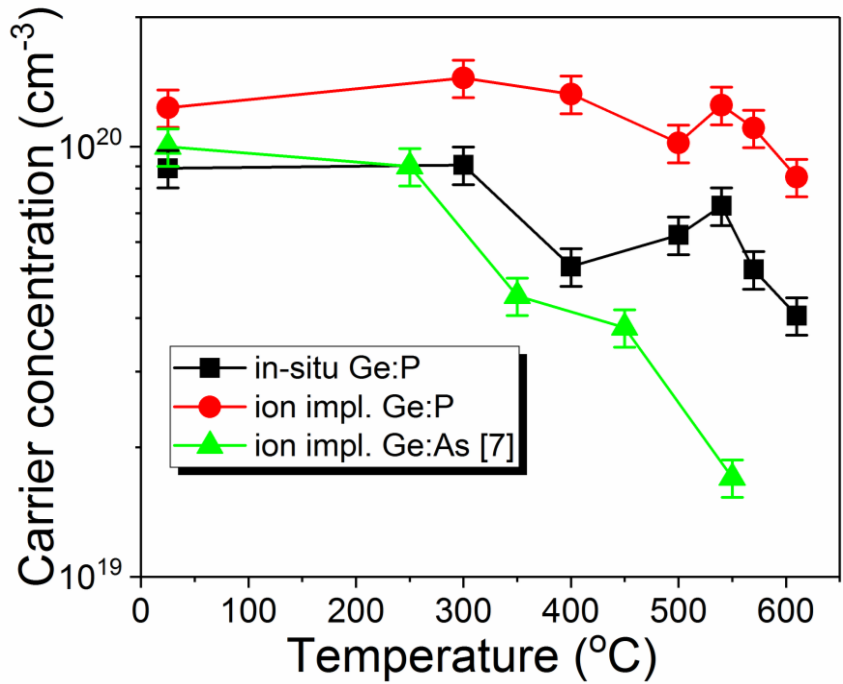

Figure 3. Thermal stability of charge carriers in heavily doped n-type Ge during post-growth annealing. The data for As doped sample is taken from Ref. 7. P doped samples were either in-situ doped or ion implanted followed by r-FLA for 20 ms at 100 Jcm$^{-2}$. The carrier concentration was estimated at RT from Hall effect measurements. P doped samples were annealed for 100 s while the As doped Ge was annealed for 10 min.

To investigate the role of the vacancy-containing clusters in more detail, PALS measurements were performed.

### 3.2. PALS measurements

Using DB-PAS in heavily P-doped Ge with a P concentration in the range of $10^{20}$ cm$^{-3}$ and an electron concentration of about $2.7\times10^{19}$ cm$^{-3}$, Kujala et al. and Vohra et al. have shown that the main defect centers which deactivate electrically active P atoms are four donors bonded with a monovacancy (P$_4$V) or with a divacancy (P$_4$V$_2$) [36, 37]. These authors analyzed Ge crystals and epilayers with a doping level and a carrier concentration very close to those in the present work. Our as-grown samples contain a nearly constant concentration of the P$_4$V clusters within the Ge layer, which is further supported by PALS measurements. The experimentally obtained positron annihilation lifetime in the as-grown sample and in the samples after FLA (1050 K for 20 ms) consists of two components $\tau_1$ and $\tau_2$. The $\tau_1$ component characterizes the annihilation lifetime of positrons within the layer, while the $\tau_2$ component corresponds to positrons annihilated at the surface state defects and vacancy clusters.

Figures 4a shows the experimentally obtained depth/energy distribution of $\tau_1$. The best fit of experimental data was achieved by fixing $\tau_2$ at 500 ps, which is the annihilation time of positrons at the surface state. The large visibility of the surface state for low positron implantation energies, $E_p < 6$ keV, is likely due to positron back diffusion to the surface. The monotonic decrease of the relative intensity $I_2$ ($I_2=100\%-I_1$) associated with $\tau_2$ confirms that in addition. The positron annihilation lifetime in intrinsic bulk Ge is $\tau=227.9$ ps, while the shortest $\tau_1$ experimentally determined for heavily P-doped Ge are in the range of 263-288 ps (at depth range of 120-300 nm) for implanted layer and about 277-298 ps for in-situ doped after FLA. This observation indicates the presence of open volume defects such as divacancies, vacancy clusters and donor atoms bonded with vacancies since all these defects lead to an increase of $\tau$ [38-40]. In the following discussion, we use the hypothesis that the lifetime component $\tau_1$ is related to $P_xV$ and $PV_2$ defects as well as divacancies.

Table 1. Calculated positron lifetimes for undoped bulk germanium and vacancy clusters without donors.

| Nr. of vacancies | 0 (bulk) | 1 | 2 | 3 | 4 | 5 | 6 | 7 | 8 | 9 | 10 |
|---|---|---|---|---|---|---|---|---|---|---|---|
| Lifetime (ps) | 227.9 | 257.8 | 326.1 | 364.5 | 401.8 | 444.6 | 486.1 | 533.7 | 571.5 | 615.1 | 660.5 |

*Ab-initio* calculations yield a positron lifetime of around 275 ps and 329 ps for $P_4V$ and $PV_2$, respectively. However, the relaxation of atoms around vacancies has not been taken into account, which causes underestimation of the calculated annihilation lifetime for donor-defect clusters of around 15 ps [36, 42, 43]. The calculated positron lifetime for neutral V and $V_2$ are found to be 260 and 320 ps [36], respectively, whereas the lifetime for large vacancy clusters in Ge is significantly higher, in the range of 400 – 800 ps, (see Table 1) [44]. The calculated positron lifetime for mono- and divacancies decorated with P atoms is summarized in Table 2. Theoretical calculations of positron lifetimes for the delocalized states (bulk lifetime) and the localized trapped states at defects were obtained by the atomic superposition (ATSUP) approach using two-component density functional theory (DFT) ab-initio calculations [41]. For the electron-positron correlation, the generalized gradient approximation (GGA) scheme is used [42]. The calculated positron lifetime for vacancy clusters in undoped Ge are summarized in

Table 1. In comparison to the neutral vacancies, inward relaxation of atoms around the vacancy is expected for charged defects, i.e. $V^{2-}$ or $V_2^{2}$, which leads to a smaller open-volume and shorter positron lifetimes [37].

Table 2. Calculated positron lifetime for mono- and divacancies decorated with up to four P atoms.

| Number of P neighbors | 0 | 1 | 2 | 3 | 4 |
|---|---|---|---|---|---|
| monovacancy lifetime (ps) | 257.8 | 262.5 | 266.9 | 271.1 | 275.1 |
| di-vacancy lifetime (ps) | 326.1 | 328.9 | 331.8 | 334.5 | 337.3 |

Hence, the calculated lifetime values are overestimated. Slotte *et al.* have shown that above 200 K neutral monovacancies in Ge become mobile tending to form stable divacancies at room temperature [40]. Therefore, the annihilation of positrons in isolated monovacancies at room temperature in heavily n-type doped Ge can be excluded. In general, the positron annihilation lifetime increases with increasing size of the open volume defect as well as with increasing number of donors bonded with vacancies, but $\tau$ slightly decreases with increasing ionization state of the defect center [45]. Importantly, all defects are in their most negative charge state when the electron concentration increases above a certain level, which significantly increases the cross-section for positron annihilation [37]. In all samples, the $\tau_1$ in the surface region (first 20 nm) is much longer than that measured in the middle of samples due to the influence of surface state defects. The measured $\tau_1$ at the medium depth for in-situ doped as-grown sample was found to be in the order of 302 ps. This is shorter than the annihilation lifetime in $V_2$, which indicates the positrons annihilation mainly in $P_4V$ centers [40] but still a certain fractions is attracted by $V_2$. At larger thicknesses positron lifetime approaches the calculated values but the overall signal superimposes with the increasing number of positrons annihilating in the substrate due to a broader depth distribution (at $E_p$=10 keV a small fraction of positrons can reach depth of even 800 nm), hence lowering artificially the lifetime value. Obtained $\tau_1$ agrees well with theoretically calculated value for $P_4V$ after considering the overestimation of the annihilation time due to neglecting the matrix relaxation, which is about 15 ps for monovacancies decorated with donors. After FLA for 20 ms, made independently from the rear-side or from the front-side, the $\tau_1$ decreases down to 294 ps at depth of ~120 nm (277 ps at 300 nm). The reduction of $\tau_1$ by 8 ps is due to reduction of the number of P atoms bonded with monovacancy by two.

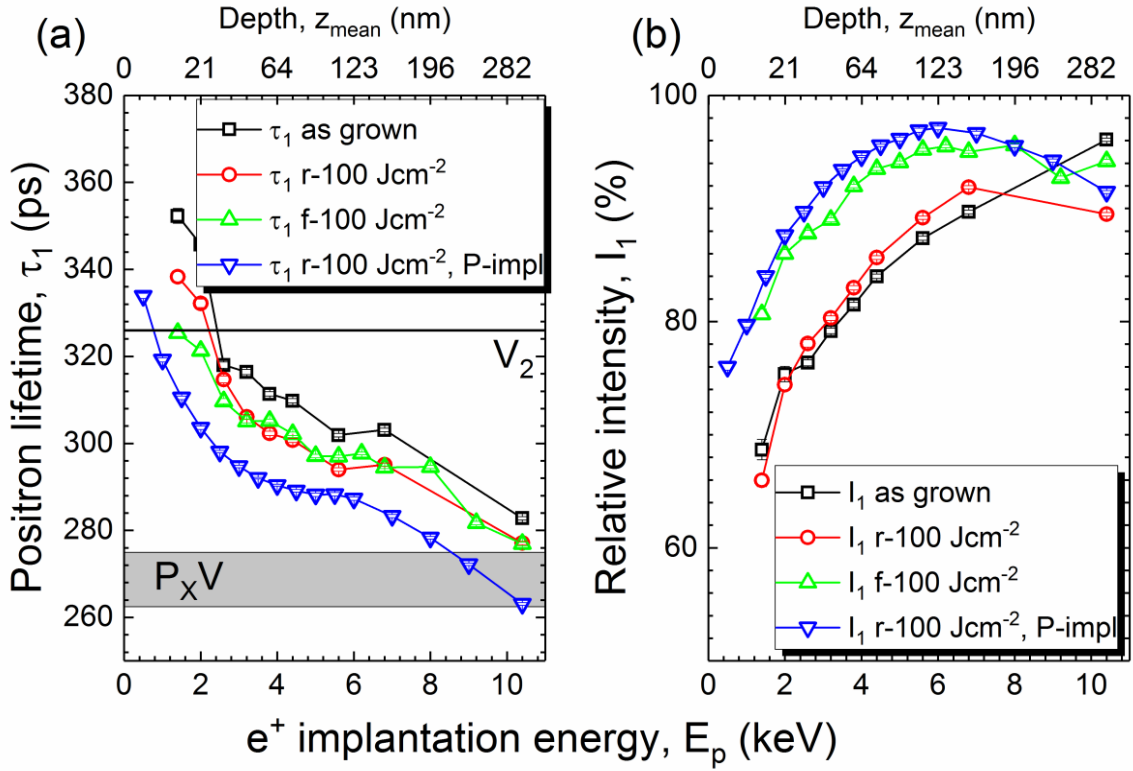

Figure 4. The positron annihilation lifetimes $\tau_1$ (a), and the intensities $I_{1,2}$ (b) as a function of the positron energy/annihilation depth obtained from as-grown and flash-lamp annealed samples and from P-implanted sample annealed from rear-side. The horizontal lines in (a) indicate the positron annihilation time in $P_x$-monovacancy and in $V_2$ taken from Table 1.

The shortest positron annihilation lifetime $\tau_1$ =286 ps (263 ps) at depth of ~120 nm (~300 nm) was measured for P-implanted sample and annealed from the rear-side. Here, we did not analyze samples implanted and annealed from the front-side because the front-side FLA of implanted Ge cause the formation of a polycrystalline layer, which stands for 1/3 of implanted thickness [28]. The $\tau_1$ measured in P-implanted and annealed Ge is shorter from the in-situ doped as-grown sample by 16 ps. Such reduction of positron annihilation lifetime suggests that the main annihilation center in P-implanted and annealed sample must be monovacancy bonded with one P atom.

In principle, before annealing, we have two different cases: (i) heavily doped epitaxial layer mainly with $P_4V$ centers which concentration depends on the doping level and growth temperature and (ii) heavily doped amorphous Ge layer with P concentration in the order of $1.8\times10^{20}$ cm$^{-3}$. In in-situ doped epitaxial layer the post-grown millisecond range non-equilibrium thermal annealing dissolves the P$_x$V clusters into isolated P atoms and monovacancies. Since the carrier concentration after FLA increases by a factor of about three, the concentration of vacancy centers bonded with P must be reduced much below $10^{19}$ cm$^{-3}$. The monovacancies in Ge are not stable at room temperature. Due to very short annealing time, only a part of them liberated from P$_4$V clusters can be trapped directly by existing larger open volume defects. From not constrained fits (not shown) we obtain lifetimes in the range of 500-800 ps with intensity below 15%, corresponding to mixture of surface states and larger vacancy clusters. During sample cooling (just after the FLA), the rest of the released monovacancies tends to form either divacancies or can be again bonded with P atoms forming P$_x$V with x<4. Due to ion implantation, on the other hand, the implanted layer becomes amorphous and during the FLA process it regrow epitaxially via explosive solid-phase recrystallization [30]. This changes significantly the final defect microstructure. The concentration of vacancies in an epitaxial layer strongly depends on the growth condition. As the growth temperature increases, the concentration of defects decreases but simultaneously the high-temperature growth ignites the dopant diffusion and segregation. Therefore, the growth conditions of doped epitaxial layers are compromises between the layer quality and doping level. During r-FLA of the sample with implanted P the temperature is close to the melting point of Ge. Therefore, the concentration of vacancies in the implanted layer recrystallized via explosive solid phase epitaxy is expected to be much lower compared to in-situ heavily-doped epitaxial layers grown by CVD or MBE methods. Simultaneously, during explosive solid phase epitaxy dopants like P are incorporated into the substitutional position of Ge and diffusion of P is suppressed. Here, for the P concentration $1.8\times10^{20}$ cm$^{-3}$ the electron concentration estimated from Hall effect is in the order of $1.2\times10^{20}$ cm$^{-3}$. The P-implanted Ge is also more stable during second annealing step. After annealing at 610 °C the effective carrier concentration is reduced by 30 % only. While the carrier concertation in in-situ P-doped or in ion implanted As-doped Ge is reduced by 55 % and 85 %, respectively. The higher stability of the ion implanted and flash lamp annealed samples is the most probably due to lower concentration of vacancies in the ultra-doped Ge-layer.

Figure 4b shows the relative intensities $I_1$ as a function of the annihilation depth/positron energy obtained from the as-grown, annealed and from implanted samples. After ion implantation and annealing, $I_1$ that is related to the concentration of positrons annihilated with lifetime $\tau_1$ is close to 100%. It indicates that positrons in the implanted and annealed sample annihilates mostly in one type of defects centers, mainly in monovacancy bonded with one P atom. The remaining positrons traps consist of surface states and vacancy clusters. The $I_1$ in epitaxial layer is in the order of ~90 %. This suggests coexistence of different positron trapping centers with much longer annihilation lifetime e.g. structural defects (vacancy clusters) like threading dislocations visible in Fig. 2. It is worth mentioning that electrically active single P atoms incorporated into the Ge lattice are positively charged (donors), which is why they cannot trap positrons. After annealing $I_1$ clearly increases reflecting overall decrease of open volume defects related to larger vacancy complexes, especially in the closer to surface sample region.

### 3.3. Theoretical calculation

For the sake of better understanding of the behavior of phosphorus-vacancy clusters as well as vacancy clusters during FLA, their thermal stability was calculated using the so-called mass action analysis [47, 48]. The following relevant clusters were considered: $P_nV$ (n=1-4), $PV_2$, and $V_n$ (n=2-4). DFT data for the binding energy of the clusters were taken from the works of Chroneos *et al.* [21] and Sueoka *et al.* [48]. Other important input parameters for the calculations were the total P concentration of $1\times10^{20}$ cm$^{-3}$ (see Fig. 1) and the total vacancy concentration of about $2\times10^{19}$ cm$^{-3}$. The latter value was estimated from the finding of Kujala *et al.* and Vohra *et al.* [36, 37] which implies that the $P_4V$ cluster is the dominating phosphorus-containing defect center in the as-grown P-doped Ge layers. This defect was assumed to contain the most part of vacancies. Figure 5 shows the calculated concentrations of the different clusters, the monovacancy and the electrically active P monomer, as a function of temperature. The results clearly confirm the assumption regarding the total vacancy concentration, since $P_4V$ is the most stable cluster that starts dissolving above a temperature of 750 K. At the deposition temperature of the heavily P-doped Ge layer (773 K), the concentration of $P_4V$ clusters and P monomers is $1.7\times10^{19}$ and $2.6\times10^{19}$ cm$^{-3}$, respectively. At about 1050 K, the concentration of $P_4V$ decreases to about $2.9\times10^{18}$ cm$^{-3}$ and that of P monomers increases up to $8.3\times10^{19}$ cm$^{-3}$. The concentration of the other clusters is below $2.5\times10^{18}$ cm$^{-3}$. Simultaneously with the dissociation of $P_4V$, the amount of monovacancies and $V_4$ increases significantly. The results of the mass action analysis are sufficient to interpret the experimental data based on the thermal equilibrium between the monomers (P and V) and the clusters. In the experiments, the electrical activation increases from about $3.5\times10^{19}$ at 773 K to values between $6.0\times10^{19}$ and $8.2\times10^{19}$ cm$^{-3}$ at 1050 K, while the calculation yields $2.9\times10^{19}$ and $8.3\times10^{19}$ cm$^{-3}$ at the respective temperatures. The small differences might be due to neglecting some of the clusters (e.g. $V_8$) in the calculation. In addition, the estimated value of the total vacancy concentration might not be completely correct. On the other hand, in the above comparison of the data from Fig. 5 with the experimental results, it is assumed that the state created in the sample at 1050 K is "frozen" and can be analyzed *ex-situ* by ECV and PALS at room temperature.

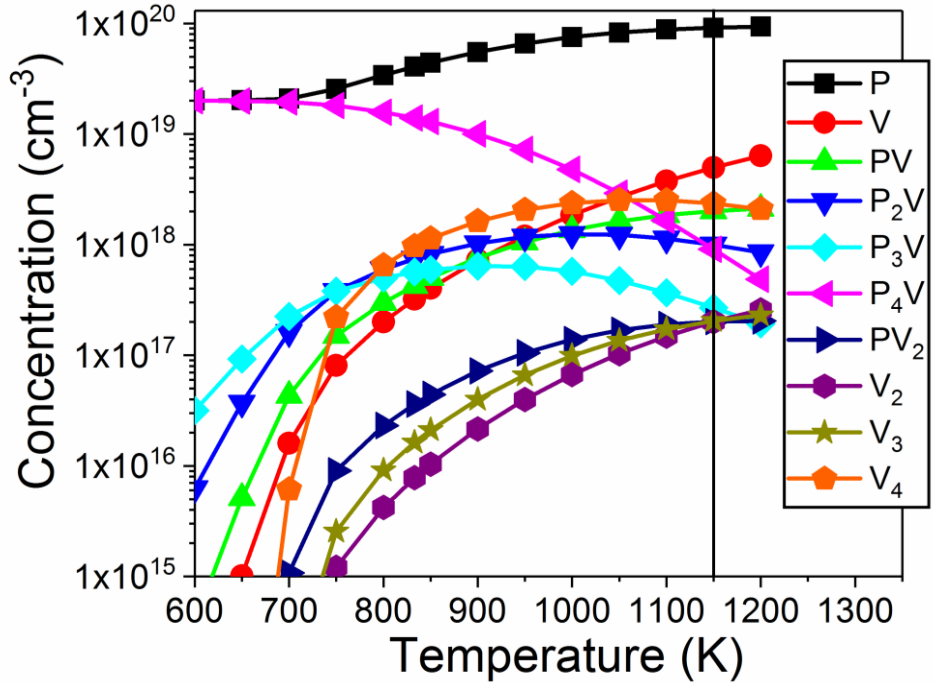

Figure 5. Calculated concentration of electrically active P and of $P_nV$, $PV_2$ and $V_n$ clusters in Ge. The total P concentration is $1\times10^{20}$cm$^{-3}$, and a total vacancy concentration of $2\times10^{19}$cm$^{-3}$ is assumed. The straight line indicates the annealing temperature.

This is certainly not completely correct, although the cooling of the sample proceeds relatively fast after FLA. This process only slightly changes the sample state, in particular the concentration of electrically active P monomers. It should be further noticed that in the mass action analysis not only thermal equilibrium but a homogeneous distribution of phosphorus and vacancies is also assumed. Both conditions are not completely fulfilled in the experiment. Nevertheless, the hypothesis used to explain the PALS results of Figs. 4 is reasonably consistent with the data depicted in Fig. 5.

Ultimately, the results of this work on P activation by FLA in in-situ doped epitaxial Ge layers are compared to those about diffusion-doping of Ge. Experimental results from SIMS and scanning spreading resistance techniques on the diffusion of n-type dopants in Ge revealed a donor (D) deactivation via the formation of neutral $D_2V$ complexes [18, 19]. The formation of this defect complex occurs during dopant diffusion at elevated temperatures up to 1025 K. Certainly, upon cooling from the temperature at which the diffusion is activated even more stable defect clusters may form via the interaction of mobile DV pairs with $D_2V$ complexes. However, since the cooling time is usually much shorter than the diffusion time, the formation of such defect clusters is effectively suppressed such that the distribution of the respective P-correlated defects formed during diffusion remains essentially unchanged. Considering these results, a FLA annealing at 1050 K is not expected to dissolve $P_2V$ complexes, which also supports the temperature dependence of the $P_2V$ concentration deduced from the mass action analysis and shown in Fig. 4. Accordingly, the remaining deactivation of P in the epitaxial Ge layer after FLA (see Fig. 1) is due to neutral $P_2V$ complexes that first dissociate at even higher temperatures. P activation by FLA of in-situ doped epitaxial Ge layers should therefore be caused by the dissolution of other complexes such as $P_4V$ as discussed above.

4. Conclusions

In conclusion, we have developed an industry-relevant method to increase the electrical activation in a heavily n-doped germanium layer. During millisecond FLA electrically active P atoms and free vacancies are released due to the dissolution of phosphorus-vacancy clusters, while the diffusion of the P atoms is effectively suppressed. Most of the liberated vacancies are trapped by large defect centers or tend to form small vacancy clusters. The remaining deactivation of P, which becomes visible by comparing the chemical and electrical dopant profiles after FLA (see Fig. 1), is likely caused by neutral $P_2V$ complexes that first dissociate at temperatures above 1050 K.

**Acknowledgement**

Support by the Ion Beam Center (IBC) at HZDR is gratefully acknowledged. The authors would like to thank R. Aniol, H. Hilliges and B. Scheumann from HZDR for their careful sample preparation. Y. Berencén would like to thank the Alexander-von-Humboldt foundation for providing a postdoctoral fellowship. Authors are grateful to the ELBE accelerator crew, who took care of delivering a stable and reliable beam. We acknowledge Federal Ministry of Education and Research (BMBF) for the PosiAnalyse (05K2013) grant, the Impulse- und Networking fund of the Helmholtz-Association (FKZ VH-VI-442 Memriox), BMBF "ForMikro": Group IV heterostructures for high performance nanoelectronic devices (SiGeSn NanoFETs) (Project-ID: 16ES1075) and the Helmholtz Energy Materials Characterization Platform (03ET7015).

**Author contribution**

S.P. writing manuscript and FLA process, M.O.L, M.B, E.H and A.W. responsible for PALS, X.W. and M.P. made theoretical calculations, J.K. and H.W. responsible for ECV

measurements, E.N. made SIMS, J.F, A.B. G.I. in-situ doping, R.H. responsible for TEM, L.R. H.B. M.H. and S.Z. data analysis and discussion.